\documentclass[onefignum,onetabnum]{siamonline171218}

\usepackage{lipsum}
\usepackage{amsfonts}
\usepackage{amsopn}
\usepackage{bbm}
\usepackage{caption}
\usepackage{subcaption}
\usepackage{graphicx}
\usepackage{epstopdf}
\usepackage{algorithmic}
\usepackage[T1]{fontenc}
\usepackage{libertine}
\usepackage{xcolor}
\usepackage{framed}
\ifpdf
  \DeclareGraphicsExtensions{.eps,.pdf,.png,.jpg}
\else
  \DeclareGraphicsExtensions{.eps}
\fi

\newcommand*\openquote{\makebox(25,-22){\scalebox{5}{``}}}
\newcommand*\closequote{\makebox(25,-22){\scalebox{5}{''}}}
\colorlet{shadecolor}{white}

\makeatletter
\newif\if@right
\def\shadequote{\@righttrue\shadequote@i}
\def\shadequote@i{\begin{snugshade}\begin{quote}\openquote}
\def\endshadequote{%
  \if@right\hfill\fi\closequote\end{quote}\end{snugshade}}
\@namedef{shadequote*}{\@rightfalse\shadequote@i}
\@namedef{endshadequote*}{\endshadequote}
\makeatother

\usepackage{enumitem}
\setlist[enumerate]{leftmargin=.5in}
\setlist[itemize]{leftmargin=.5in}

\newsiamremark{remark}{Remark}
\newsiamremark{hypothesis}{Hypothesis}
\crefname{hypothesis}{Hypothesis}{Hypotheses}
\newsiamthm{claim}{Claim}

\headers{Deep RL with Function Properties in Mean Reversion Strategies}{Sophia Gu}

\title{Deep Reinforcement Learning with Function Properties in Mean Reversion Strategies}

\author{Sophia Gu\footnote{Department of Mathematics, Courant Institute of Mathematical Sciences, New York University.\\
E-mail: \email{keying@nyu.edu}}}

\ifpdf
\hypersetup{
  pdftitle={Deep Reinforcement Learning with Function Properties in Mean Reversion Strategies},
  pdfauthor={Sophia Gu}
}
\fi

\begin{document}
\maketitle

\begin{abstract}
  Over the past decades, researchers have been pushing the limits of Deep Reinforcement Learning (DRL). Although DRL has attracted substantial interest from practitioners, many are blocked by having to search through a plethora of available methodologies that are seemingly alike, while others are still building RL agents from scratch based on classical theories. To address the aforementioned gaps in adopting the latest DRL methods, I am particularly interested in testing out if any of the recent technology developed by the leads in the field can be readily applied to a class of optimal trading problems.\\
  Unsurprisingly, many prominent breakthroughs in DRL are investigated and tested on strategic games -- from AlphaGo to AlphaStar and at about the same time, OpenAI Five. Thus, in this writing, I want to show precisely how to use a DRL library that is initially built for games in a fundamental trading problem -- mean reversion. And by introducing a framework that incorporates economically-motivated function properties, I also demonstrate, through the library, a highly-performant and convergent DRL solution to decision-making financial problems in general.
\end{abstract}
\bigskip
\section{Introduction}
Mean reversion strategy has been studied for decades. In 1994 \cite{La94} first looked at mean reversion in earnings and by contrasting past and future growth rates, they found that earnings tend to regress to their historical mean over time. Following that, many mean reversion strategies have emerged. In 2008 \cite{AvLe08} built a mean reversion model using trading signals generated from PCA and sector ETFs, and related the performance of mean reversion statistical arbitrage with the stock market cycle. Other authors, like \cite{Be09} and \cite{LiPr20}, have attempted to derive analytical solutions for optimal trades when the underlying price follows an Ornstein-Uhlenbeck (OU) process. While those are ground-breaking findings, they only admit explicit solutions in the specific cases being studied and are often vulnerable to numerical errors in practice. More recent mean reversion strategies started to make use of growing computing power as well as Reinforcement Learning (RL) algorithms, a younger sibling to Optimal Control introduced in \cite{Be57}. For example, \cite{Ri17} used a classical RL algorithm -- tabular Q-learning -- for simple problems approximated with an OU-driven price process and achieved an average Sharpe ratio close to $2.07$.

One advantage of DRL compared to tabular Q-learning is that it permits continuous state and action spaces, which is often the limiting factor that prevents a trained model from achieving its theoretical optimal performance. Although due to time limitation, I could only evaluate my agent's performance on mean reversion problems, the same model can be easily reconfigured for other stochastic systems that admit arbitrage opportunities. For instance, a single-factor APT model has also been briefly studied -- it indeed yielded promising results similar to mean reversion\footnote{Code is available in the accompanying GitHub repository.}.

That said, the landscape of a DRL policy function is often so complex that in most situations, an agent searching for a good local optimum is like navigating a path covered in fog; it would so easily get stuck in a small puddle before reaching anywhere close to its goal. However in finance, investors often possess insights of optimal policies from either economic principles or insider knowledge. As an example, let's take a closer look at mean-reverting price processes.

Due to short-term pricing inefficiencies, a price process in stock market can exhibit mean reversion property, which implies a near arbitrage in the system; when the price is too far out of its equilibrium, a trade betting that it returns back to the equilibrium has a slim chance of loss. In an RL context, it is equivalent to say that we expect the action suggested by the policy network to be monotonically decreasing \textit{w.r.t.} price. So although one does not know the exact optimal solution to a trading problem, it is often not that difficult to come up with shapes that govern it. Surprisingly, this information is often absent when training RL agents, but turns out to be critical as it effectively scopes the target function space and thus, to some extent, solves the complex landscape problem that I mentioned at the beginning.
\bigskip

This article is organized as follows: \cref{sec:preliminary} lists all the necessary preliminaries for understanding the study -- it introduces utility theory as well as the very algorithm that my chosen software is based upon on; \cref{sec:framework1} and \cref{sec:framework2} are the main course where I lay out the theoretical framework for DRL with function properties, which is then followed by introducing the software and the neural net architecture in \cref{sec:setup}; \cref{sec:experiments} enumerates the results I obtained on two representative mean-reverting price processes -- an OU process and a more general ARMA process.

The data supporting the findings of this study is available in the GitHub repository: \url{https://github.com/sophiagu/RLF}.
\bigskip

\section{Preliminaries}
\label{sec:preliminary}
\subsection{Utility theory}
\label{sec:utility}
The reward function I use is based on utility theory formalized by \cite{Pr64} and \cite{Ar71}. Under their framework, suppose a rational investor invests in a stock over a finite time horizon: 1, 2, ..., $T$. She then chooses actions to maximize the expected utility of her terminal wealth
\begin{displaymath}
    \mathop{\mathbb{E}}[u(w_T)] = \mathop{\mathbb{E}}[u(w_0 + \sum_{t=1}^T \delta w_t)]
\end{displaymath}
where $w_0$ is the initial wealth and $\delta w_t = w_t - w_{t-1}$ is the change in wealth at each timestep. The function $u: \mathop{\mathbb{R}} \rightarrow \mathop{\mathbb{R}}$ denotes the utility, which is a mapping from wealth to a real number. In order for the utility to make sense in real world, $u$ should be increasing. In addition, assume the investor is risk-averse, then $u$ is also concave. Refer to \cite{Pr64} and \cite{In87} for more details on the shape of $u$.

It turns out that we can simplify the above expectation by assuming the return of the stock follows a \textit{mean-variance distribution}.

\begin{definition}[Mean-Variance Distribution]\label{thm:meanvar}
  The underlying random asset return $r$ is said to follow a mean-variance equivalent distribution if it has a density $\rho(r)$, has first and second moments, and for any increasing utility function $u$, there exists a constant $\kappa > 0$ such that the policy, $\pi^{\star}$,  which maximizes $\mathop{\mathbb{E}}[u(w_T)]$ is also optimal for the simpler problem
  $$\max_\pi \{\mathop{\mathbb{E}}[w_T] - \frac{\kappa}{2}\mathop{\mathbb{V}}[w_T]\}$$
\end{definition}

 By writing $w_T = w_0 + \sum_{t=1}^T \delta w_t$, the expected utility can be further reduced to
\begin{displaymath}
    \max_\pi \sum_t \{\mathop{\mathbb{E}}[\delta w_t] - \frac{\kappa}{2}\mathop{\mathbb{V}}[\delta w_t]\}
\end{displaymath}

\subsection{Deep reinforcement learning}
The basic RL problem consists of an environment and an agent. At each iteration, the agent observes the current state of the environment and proposes an action based on a policy. After each interaction, the agent receives a reward from the environment and the environment updates its state. The change in the environment state can be autonomous (\textit{e.g.} the stochastic evolution of a stock's price), or can be influenced by the agent's action (\textit{e.g.} the change in the bid offer spread after the agent exercises a trade). DRL differs from RL in that it trains a neural network to learn that policy\footnote{In this writing, I call such a neural net a policy network.}. \textit{Fig. 1} compares RL and DRL in a pictorial way.

\begin{figure}[htbp]
\centering
  \begin{subfigure}{\textwidth}
    \centering
    \includegraphics[width=80mm,scale=1]{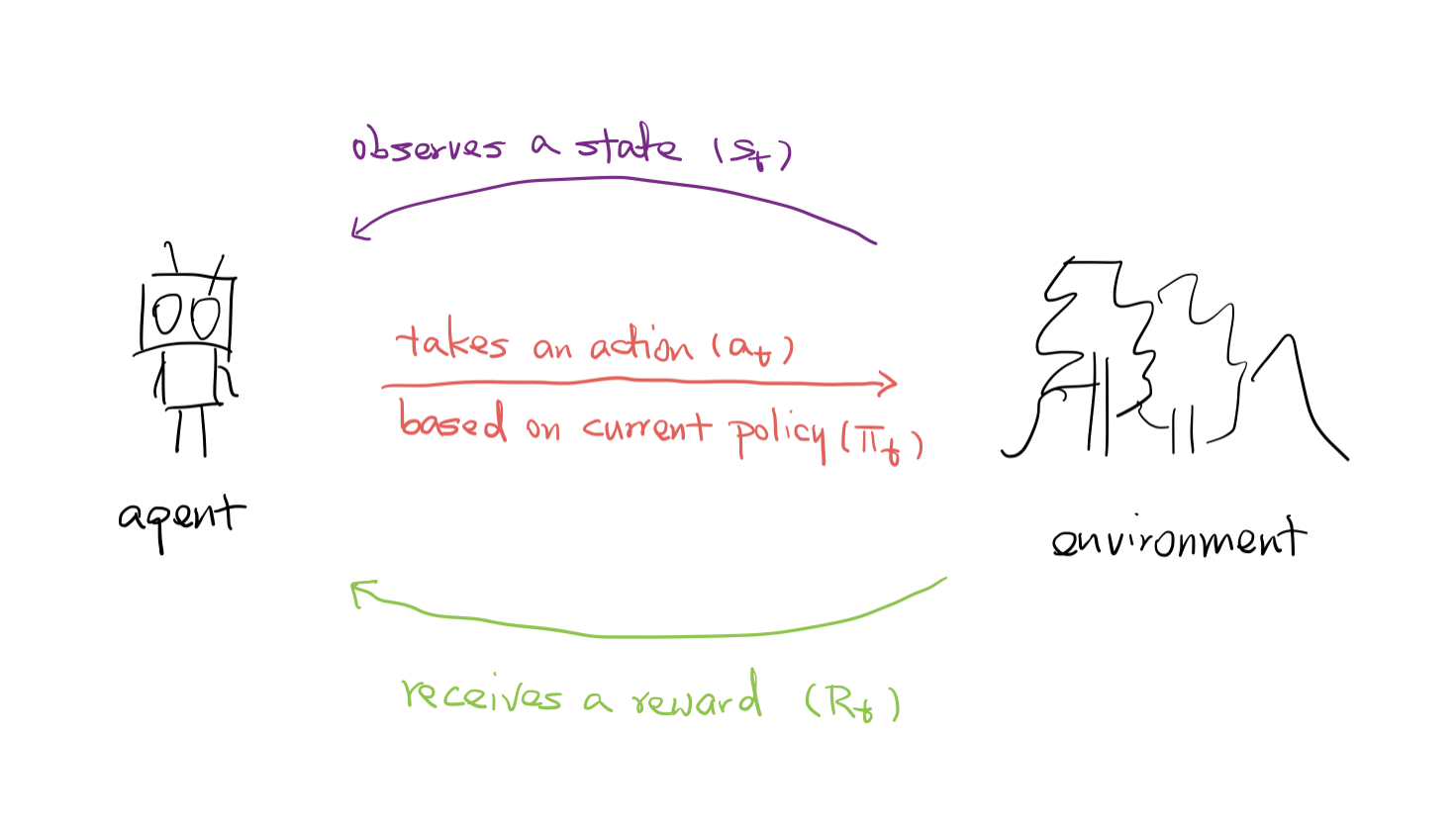}
  \end{subfigure}
  \begin{subfigure}{\textwidth}
    \centering
    \includegraphics[width=90mm,scale=1]{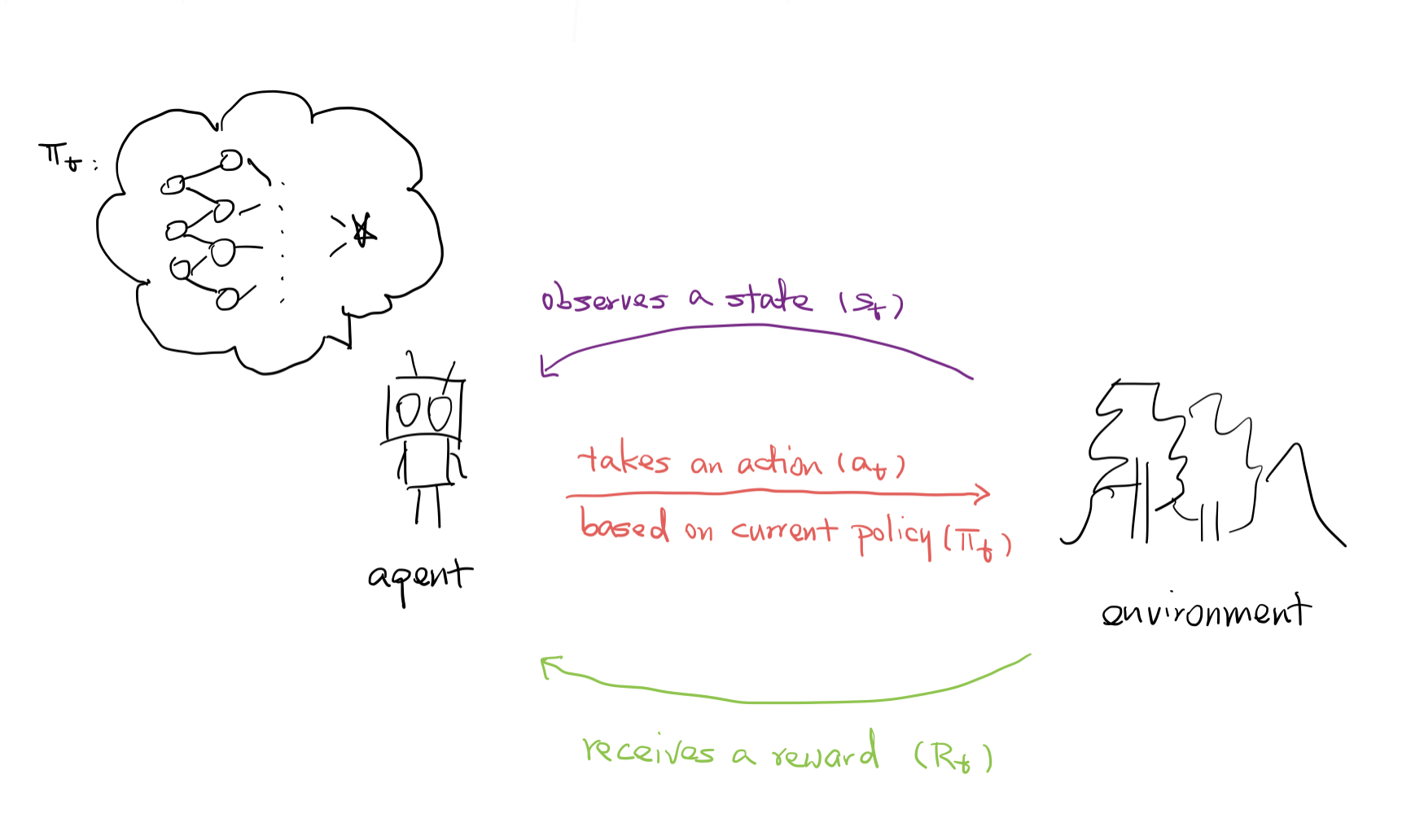}
  \end{subfigure}
  \caption{One iteration of RL (\textit{top}) and DRL (\textit{bottom}) procedure}
  \label{fig:rl1fig}
\end{figure}

\subsection{Proximal Policy Optimization}
When comes to picking a specific RL algorithm, I focused on model-free algorithms as they have been studied more extensively than model-based algorithms. Within the scope of model-free RL algorithms, the two big branches are Q-learning and policy optimization. While I have experimented with both approaches, policy optimization yielded more promising results given the same amount of training time. For policy optimization, besides its built-in support for continuous state space and action space, it directly improves the policy as it updates the policy network by following the gradients \textit{w.r.t.} the policy itself; whereas in Q-learning the updates are done based on the estimates of the value function, which only implicitly improves the policy. It turns out that Q-learning also tends to be less stable than policy optimization algorithms. For references, see \cite{TsBe97}, \cite{Sz09} and Chapter 11 of \cite{SuBa18}.

As a result, I settled down to a policy gradient based DRL algorithm called Proximal Policy Optimization (PPO). PPO is one of the Actor-Critic algorithms that have two neural networks, one for estimating the policy (actor) function and the other for the value (critic) function. The main policy gradient loss function is
\begin{displaymath}
    L^{PG}(\theta) = \hat{\mathop{\mathbb{E}}}_t[\log \pi_\theta(a_t|s_t)\hat{A}_t]
\end{displaymath}

Here, $\theta$ represents the parameters or weights of a neural net, $\pi_\theta(a_t|s_t)$ is the probability of choosing an action $a_t$ based on a state $s_t$, and $\hat{A}_t$ is an estimate of the advantage function, \textit{i.e.}, the relative value of the selected action to the base action.

To fully understand this equation and its subsequent variations, I refer to the well-written original paper on PPO\cite{Sc17}. Intuitively, this loss function tells an agent to put more weight on a good policy. In other words, to assign higher probabilities to actions that lead to higher critic values and vice versa.

Another thing to keep in mind is that PPO combines ideas from A2C (having multiple workers) and TRPO (which uses a trust region to improve the actor). The gist is, in order to avoid overfitting, the new policy should be not too far from the old one after an update. For that, PPO uses clipping to avoid too large updates. Nevertheless, it is helpful to look at the training loop to understand how the agent learns.

\begin{algorithm}
\caption{PPO, Actor-Critic Style}
\begin{algorithmic}
\FOR{iteration := 1, 2, ...}
\FOR{actor := 1, 2, ..., $N$}
\STATE{Run policy $\pi_{\theta old}$ in environment for $T$ timesteps}
\STATE{Compute advantage estimates $\hat{A}_1$, ..., $\hat{A}_T$}
\ENDFOR
\STATE{Optimize surrogate $L$ w.r.t. $\theta$, with $K$ epochs and minibatch size $M\le NT$}
\STATE{$\theta_{old} := \theta$}
\ENDFOR
\end{algorithmic}
\end{algorithm}
\bigskip

\section{Deep reinforcement learning setup}
\label{sec:framework1}

In this section, I break down the RL setup into three main components -- a state space, an action space and a reward function. I will not labor over the bulk of derivations here since they are already well-known in the field. If you are not familiar with the framework though and want to see some rigorous derivations, one good reference is \cite{Ri17}.

\subsection{State space}
The term \textit{state}, in RL, refers to the state of the environment; it is a data structure consisting of all the information that an agent needs for choosing an action. I use $s$ to denote the state of the environment and index it by time $t$, meaning it is the information available at that particular timestep. For mean reversion in stock market, clearly we will need the current holding of the stock, $h_t$, and its current price, $p_t$. Besides, I also include the stock price from an earlier timestep, $p_{t-1}$. Recall that underlying every RL problem is a Markov Decision Process, which means that an agent should be able to make a decision at timestep $t$ solely based on the information present in $s_t$ rather than any information in the prior states, $s_{\le t}$. Including the earlier price, which tells the agent not only the current price but also the change in price, makes it possible for it to learn the underlying price dynamics.

\subsection{Action space}
If we are building a trading agent to try to hedge or to maximize a reward function, then the action space is typically the trade that the agent does. For mean reversion, this is equal to $h_t - h_{t-1}$ at timestep $t$.

\subsection{Reward function}
The goal of RL is to maximize a cumulative reward (or utility), or to minimize a cumulative cost over time. Without loss of generality, since one is just the negation of the other, in this article we will only discuss reward maximization. Following the discussion in \textit{Section \ref{sec:utility}}, we can approximate the reward function at each timestep $t$ as a mean-variance equation
\begin{displaymath}
    R_t \approx \delta w_t - \frac{\kappa}{2}(\delta w_t)^2
\end{displaymath}
where, in our problem, $\delta w_t$ is the P\&L at timestep $t$ and can be written precisely as $(p_t - p_{t-1})h_{t-1} - cost_t$. Notice we also account for transaction costs accompanying the trade.

The transaction cost model I use is rooted in \cite{GaPe13}. To compute trading cost, imagine we are looking at a limit order book: for fairly liquid securities the bid offer spread is either one tick or two ticks. In the US, a tick is a penny. Typically the displayed liquidity on the bid and offer is a few hundred shares. Let’s suppose it is one hundred shares. To buy shares, we cross the spread and take out the offer. In other words, we transact one lottery a hundred shares at the offer and we cross the spread to do it. If the spread is two ticks wide then that is the cost of one tick relative to the midpoint. Extend the idea linearly, if we trade ten lots then we move the price ten ticks.\footnote{For larger trades, that may not be a good approximation, but for small trades that is not too far off reality.}
\bigskip

\section{Function properties}
\label{sec:framework2}

Inspired by \cite{SiAb97} which introduced a monotonic hint for classification problems, in this article, I will instead focus on a DRL setting, and build a framework of incorporating more general function properties into its training process. I also give a few examples illustrating the idea of the framework; how it applies to specific problems like option pricing and statistical arbitrage -- in particular, mean reversion strategies.

\subsection{Motivation and Definitions}
When searching for an optimal trading strategy using a neural network, we are effectively looking for an optimal function in a space parametrized by
$$\mathop{\mathbb{P}} = \{all\;distributions\;on\;\mathop{\mathbb{R}}\;that\;can\;be\;expressed\;by\;a\;neural\;network\}$$

Notice that this space is huge. But Stochastic Gradient Descent (SGD) is inherently only working in a low-dimensional subspace and cannot explore the whole space of the parameters. The subspace over which SGD operates is spanned by all the stochastic gradients along the trajectory and these stochastic gradients are highly correlated with each other. Thus, the dimensionality of such a subspace is upperbounded by the number of SGD updates, which is typically a small number compared to the space itself. One way to reduce the function space is to ``encourage'' the underlying algorithm to search in a desirable subset(s) of it. This leads us to introducing \textit{function penalty}. But before that, let's first define some terms that will become useful later on.
\bigskip
\begin{shadequote}
    In fact, mathematics is, to a large extent, invention of better notations.\par\textemdash\emph{Richard Feynman}
\end{shadequote}
\bigskip
All the following definitions assume that each training stage is measured by one epoch, \textit{i.e.}, the parameters of a policy network are only updated in between two epochs.

\begin{definition}[Policy Function]\label{thm:policyfn}
  Let $\mathbb{I}$ be an input space and $\mathbb{A}$ be the corresponding action space. $a: \mathbb{I} \rightarrow \mathbb{A}$ is a policy function that takes an input $i$ and produces either a deterministic or stochastic action in $\mathbb{A}$. $a$ is usually parametrized by a set of parameters $\theta \in \Theta$, where $\Theta$ is specified by a given problem.
\end{definition}

\begin{definition}[Function Penalty]\label{thm:fnpenalty}
  Let $B$ be a subset in $\mathbb{I} \times \mathbb{A}$.
  A characteristic function $l^B: \mathbb{I} \times \mathbb{A} \rightarrow \{0,1\}$ is called a function penalty if it takes an input and policy pair $(i,a)$, outputs zero when $(i,a)\in B$ and a unit penalty otherwise.
\end{definition}

One can think of a function penalty as an indicator of whether an action suggested by a policy network satisfies a given constraint or condition, defined by the set $B$. For example, let $x\in [-1,1]$ be the $x$-coordinate of a vector in $\mathbb{R}^2$, and $y=a(x) \in [-1,1]$ be the $y$-coordinate of the vector. If say, one expects the vectors $(x,y)^T$ to lie in a unit circle centered at the origin, then she can define $$B=\{(x,y)\in [-1,1]\times [-1,1]|x^2+y^2 \le 1\}$$ which results in $$l^B(x;y) = \mathbbm{1}_{\{\neg B\}} = \mathbbm{1}_{\{x^2+y^2>1\}}$$

As you may start to see, if we sample more and more $x$s and get $y$s produced by $a(x)$, then according to the Law of Large Numbers, we will be more and more confident about our understanding of the policy's performance by summing up the values of $l^B(x;y)$. That gives the idea of the next definition.

\begin{definition}[Cumulative Function Penalty]\label{thm:fnpenalty2}
  Given a set of inputs $I=\{i\}_{j\in \mathbb{N}}$, a group of function penalties $L=\{l\}_{k\in \mathbb{N}}$ and a policy $a$, define the cumulative function penalty $f$ as
  $$f(I, L, a) = \lambda \sum_{i_j\in I} \bigvee_{l_k\in L} l_k(i_j; a(i_j))$$
  where $\lambda$ is a scalar and $\bigvee$ denotes a logical OR operation.
\end{definition}

It is important to pick $\lambda$ so that $f(I, L, a)$ is on a similar scale to the remaining part of the reward function. Before pressing on to the next topic, I do want to emphasize a few places in finance where economically meaningful constraints can be incorporated into training using this framework.

\subsection{Examples in finance}

\paragraph{Option pricing}

Consider the model-free constraints on the shape of option pricing function. For instance, the price of a European call should satisfy, assuming a compounded interest rate $r$, $$c(S, \tau, K) \ge (S-Ke^{-r\tau})_+$$
where $S$ is current stock price, $K$ is strike price and $\tau$ is time to maturity.

To see the above inequality holds, consider no-arbitrage arguments and observe that holding a long call is no worse than holding its corresponding forward contract, and thus $c(S, \tau, K) \ge S-Ke^{-r\tau}$; neither is holding a long call worse than holding nothing, which is equivalent to saying $c(S, \tau, K) \ge 0$.

A function penalty corresponding to the above constraints is simply
$$l_1(S, \tau, K; c) = \mathbbm{1}_{\{c < 0 \}} \vee \mathbbm{1}_{\{c < S-Ke^{-r\tau}\}}$$
which emits a unit penalty when either of the constraints is not satisfied. Using a similar argument, one can also show that $$c(S, \tau, K) \le Se^{-d\tau}$$
where $d$ is a compounded dividend yield\footnote{One derivation is available in \cite{BiPi09}.}. So we can introduce yet another function penalty
$$l_2(S, \tau, K; c) = \mathbbm{1}_{\{c > Se^{-d\tau} \}}$$
Combining $l_1$ and $l_2$, we get the cumulative function penalty for pricing a European call option

\begin{align*}
f(I, \{l_1, l_2\}, c) &= \lambda \sum_{S, \tau, K\in I} l_1(S, \tau, K; c) \vee l_2(S, \tau, K; c)\\
&= \lambda \sum_{S, \tau, K\in I} \mathbbm{1}_{\{c < 0 \}} \vee \mathbbm{1}_{\{c < S-Ke^{-r\tau}\}} \vee \mathbbm{1}_{\{c > Se^{-d\tau} \}}
\end{align*}

It is worth pointing out that the inputs to a function penalty need not be unary. So instead of passing one stock price into $l$, one may choose to pass several prices. For example, observe that the payoff $c(S,K) = (S-K)_+$ is a convex function in $S$. Again by no-arbitrage arguments: $$\frac{\partial^2 c}{\partial S^2}=\delta (S-K)=\rho (K)>0$$
where $\delta (\cdot)$ is the Dirac delta function and $\rho (\cdot)$ is the density function of $S$.

One way to construct a function penalty for constraining $c$ to the set of convex functions in $S$ is to directly apply its definition: \textit{A function $f(x)$ is convex iff $f(\kappa x+(1-\kappa)y) \le \kappa f(x)+(1-\kappa)f(y)$ for $\kappa \in (0,1)$.}
This gives, for two different stock prices $S_1, S_2$,
$$l(S_1, S_2; c)=\mathbbm{1}_{\{c(\kappa S_1+(1-\kappa)S_2) > \kappa c(S_1)+(1-\kappa)c(S_2)\}}$$

\paragraph{Statistical arbitrage}

Now let's look at some examples exploiting arbitrage opportunities. Recall that when a stock price has a mean reversion nature, optimal trades should be a decreasing function in price. We can, following a similar fashion, construct a cumulative function penalty for mean reversion strategies. Since we will code up this particular function penalty for the study, it deserves a special definition.

\begin{definition}[Cumulative Function Penalty for Mean Reversion]\label{thm:fnpenalty3}
  Given a set of stock prices generated for one epoch of training, $p_1,..., p_T\in \mathop{\mathbb{R}}^+$, and consider a policy network that maps a stock price to a trade, $a$. Let $i, j$ be two integers sampled uniformly from 1 to $T$ such that $i\neq j$. To simplify the notation, write $a_i=a(p_i)$, $a_j=a(p_j)$, and denote $I=\{p_i, p_j\}_{i,j\in[1,T]}$.\\
  We use the following function penalty
  $$l_{mr}(p_i, p_j; a)=\mathbbm{1}_{\{(p_i < p_j) \oplus (a_i > a_j)\}}$$
  and its corresponding cumulative function penalty for mean reversion
  $$f_{mr}(I, l_{mr}, a) = \lambda \sum_{p_i, p_j\in I} l_{mr}(p_i, p_j; a)$$
  where $\oplus$ represents a logical XOR operation.
\end{definition}

We conclude this section with a closely related mean reversion strategy -- pairs-trading. Suppose a pair of stock prices $(p_t, q_t)^T$ has a cointegration vector $(1, -\alpha)^T$ where $\alpha \in \mathbb{R}$, \textit{i.e.}, a portfolio formed by $\pi_t=p_t -\alpha q_t$ is stationary and hence mean-reverting. Pairs-trading strategy takes a long position on the portfolio when its value drops and reverts its position when the value goes up. So the trade for $\pi_t$ behaves exactly like the simple mean reversion strategy we described above. As a result, in \textit{Definition \ref{thm:fnpenalty3}}, one can simply substitute $p_i$ by $\pi_i$ to get a ``free'' cumulative function penalty for pairs-trading.

\subsection{New reward function}

In a nutshell, function penalty tells us how likely or unlikely a candidate policy function, $a$, is given a specific problem setting, $D$. In the world of Bayesian, this is called the \textit{a priori} probability density of a candidate function. Here we denote it as $\mathbb{P}(a|D)$.

Let $R$ be the original reward, then we can use Bayes's Theorem to derive the \textit{posterior} likelihood of a policy function given both the problem setting and the original reward
\begin{displaymath}
    \mathop{\mathbb{P}}(a|D,R) \propto \mathop{\mathbb{P}}(R|D,a) \mathop{\mathbb{P}}(a|D)
\end{displaymath}
Taking $log$ on both sides and plugging in the cumulative function penalty from the prior section to obtain
\begin{align*}
    \log \mathop{\mathbb{P}}(a|D,R) &\propto
    \log \mathop{\mathbb{P}}(R|D,a) + \log \mathop{\mathbb{P}}(a|D)\\
    &\propto R - f_{mr}
\end{align*}
Putting everything together, we arrive at a new reward function
\begin{displaymath}
    R^*_t \approx \delta w_t - \frac{\kappa}{2}(\delta w_t)^2- f_{mr}
\end{displaymath}
\bigskip

\section{Experimental setup}
\label{sec:setup}
\subsection{Software}
I use OpenAI's improved version of its original implementation of PPO, Stable Baselines\footnote{\url{https://github.com/hill-a/stable-baselines}}, running on TensorFlow. This release of OpenAI Baselines includes scalable, parallel implementations of PPO which uses MPI for data passing.

\subsection{Network architecture}
For both value and policy networks, I use a $64\times 64$ feedforward neural net with ReLU activation function followed by an LSTM layer with 256 cells. This network is designed to be slightly bigger than the actual size of the problem for easier optimization.\footnote{As pointed out by \cite{Ru86}: "Add a few more connections creates extra dimensions in weight-space and these dimensions create paths around the barriers that create poor local minima in the lower dimensional subspaces."} Both training and hyperparameters tuning use an Adam optimizer with a learning rate of $1e-5$ and early stopping. I counter the problem of potential overfitting by adding $l1$ and $l2$ regularizations with rates of $0.01$ and $0.05$ respectively. These regularizations encourage smaller and more sparse learned weights.

Further, each model uses 5 random restarts and auto-selects the parameters that give the best performance for out-of-sample testing.

\subsection{Simulation}
I use simulated stochastic price processes -- an OU process and an ARMA(2,1) process -- for training and testing DRL models.

Suppose that there is some equilibrium price, $p_e$, a variance $\sigma^2$, and a positive mean reversion rate $\lambda$. Let $x_t = \log(p_t/p_e)$, then the OU process has the dynamics
$$dx_t = -\lambda x_t dt + \sigma dW_t$$
where $W_t$ is a Wiener process.
And the ARMA(2,1) process has the dynamics
$$dx_t = -(\lambda_1 x_t + \lambda_2 x_{t-1}) dt + \sigma_1 dW_t + \sigma_2 dW_{t-1}$$
Both processes are stationary\footnote{You may want to use the unit root test to check them out.} and therefore mean-reverting.

For simplicity, I pick $dt=1$ in the Monte Carlo simulations. It follows that $dW_t \approx \sqrt{dt}\epsilon_t = \epsilon_t$, where $\epsilon_t \sim \mathcal{N}(0,1)$ is \textit{iid} white noise\footnote{The trajectory of $x_t$ can be sampled exactly. \textit{E.g.}, in the OU process, one can instead draw samples from $\mathcal{N}(0,\frac{\sigma^2}{2\lambda}(1-e^{-2\lambda t}))$ with arbitrary timestep $t$.}.

\subsection{Success criteria}
To compare different agents' performance, the standard is to use
\begin{align*}
    annualized\;Sharpe\;ratio &= \frac{260 \times mean\;of\;daily\;P\&Ls}{\sqrt{260} \times standard\;deviation\;of\;daily\;P\&Ls}\\
    &\approx 16\frac{mean\;of\;daily\;P\&Ls}{standard\;deviation\;of\;daily\;P\&Ls}\\
\end{align*}
along with its standard deviation, timesteps to convergence, etc.
\bigskip

\section{Results}
\label{sec:experiments}
I ran 10,000 Monte Carlo simulations for evaluating different models' out-of-sample performance using annualized Sharpe ratio for both an OU process and an ARMA process. \textit{Table 1} and \textit{Table 2} display their Sharpe ratios respectively. To make the notations easier to follow, I denote \textit{Model A} for agents trained using the original mean-variance reward function and \textit{Model B} for agents trained using the new augmented reward function. For the OU process, I also have a chance to compare my agents to the tabular Q-learning model from \cite{Ri17}.

\begin{center}
\begin{table}[htbp]
\caption{Sharpe ratios of the trades from 10,000 out-of-sample simulations of OU process}
\centering
\begin{tabular}{ |c|c|c|c| }
 \hline
 model & Q-learning \textit{(benchmark)} & Model A & Model B (\textit{Model A+fn property}) \\
 mean & 2.07 & 2.10 & 2.78 \\
 std & NA & 0.375 & 0.329 \\
 timesteps to convergence & 1000k & 7k & 4k\\
 \hline
\end{tabular}
\end{table}
\end{center}

\begin{center}
\begin{table}[htbp]
\caption{Sharpe ratios of the trades from 10,000 out-of-sample simulations of ARMA process}
\centering
\begin{tabular}{ |c|c|c| }
 \hline
 model & Model A & Model B (\textit{Model A+fn property}) \\
 mean & 2.46 & 3.22 \\
 std & 0.479 & 0.268 \\
 timesteps to convergence & 4k & 10k\\
 \hline
\end{tabular}
\end{table}
\end{center}

\textit{Fig. 2} and \textit{Fig. 3} below show the kernel density estimates of the Sharpe ratios of all simulation paths. The idea of kernel density estimates is to plot the observed samples on a line and to smooth them so that they look like a density.

\begin{figure}[htbp]
\centering
\begin{subfigure}{\textwidth}
  \centering
  \includegraphics[width=50mm,scale=1]{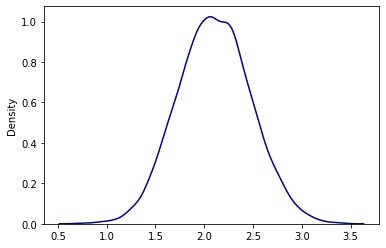}
\end{subfigure}
\begin{subfigure}{\textwidth}
  \centering
  \includegraphics[width=50mm,scale=1]{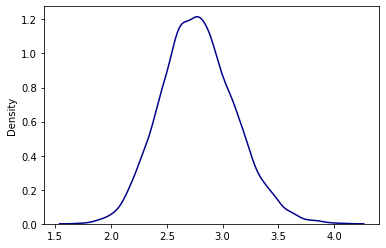}
\end{subfigure}
\caption{Kernel density estimates of the Sharpe ratios from 10,000 out-of-sample simulations of OU process: Model A (\textit{top}), Model B (\textit{bottom})}
\end{figure}

\begin{figure}[htbp]
\centering
\begin{subfigure}{\textwidth}
  \centering
  \includegraphics[width=50mm,scale=1]{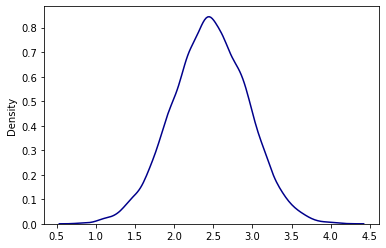}
\end{subfigure}
\begin{subfigure}{\textwidth}
  \centering
  \includegraphics[width=50mm,scale=1]{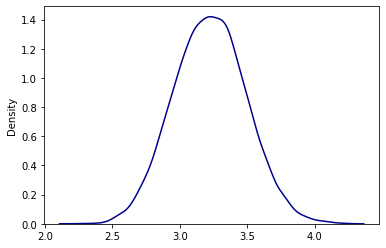}
\end{subfigure}
\caption{Kernel density estimates of the Sharpe ratios from 10,000 out-of-sample simulations of ARMA process: Model A (\textit{top}), Model B (\textit{bottom})}
\end{figure}
For both processes, we observe a noticeable increase in the average Sharpe ratio and a decrease in variance when incorporating a function property in training.
Moreover, for each process, I performed a two-sample t-test to determine if we have high confidence in the improvement in performance. The differences in Sharpe ratios are indeed highly statistically significant, with t-statistics of 135 and 139 respectively.

Another interesting observation is that with DRL, the agents were able to converge within 10 epochs, equivalent to 10,000 timesteps as I set 1,000 timesteps for each epoch, and the fastest only took 4,000 steps, much faster than the tabular Q-learning agent, which took about one million training steps.
\bigskip

\section{Conclusion}
\label{sec:conclusion}
This article demonstrates how to apply DRL to quantitative financial problems using the latest technology developed by the pioneers in the field, and how to use domain knowledge to encourage the underlying training algorithm for finding better local optima. I provide a proof of concept in a controlled numerical simulation which permits an approximate arbitrage, and I verify that the DRL agent finds and exploits this arbitrage in a highly efficient way, while producing Sharpe ratios that surpass prior state-of-the-arts.
\bigskip

Before concluding, I leave open two avenues to look into further:

1) With simple function penalties, I did not notice any increase in training time, but if you are to incorporate more convoluted properties which have high demand on computation time at each epoch, it may indeed slow down the training process. One potential solution is to build the function property directly into the network architecture\footnote{Indeed I have already made some attempts here: \url{https://github.com/sophiagu/stable-baselines-tf2/blob/master/common/policies.py}};

2) I only trained the agents in a purely simulated environment, it would be interesting to see how they perform in the real market. One foreseeable challenge is that we will not have as much real-world data as simulated data, but we can train on simulated data first to find a good initialization of the network weights before continue training the model on the limited market data.
\bigskip

\section*{Acknowledgments}
I am grateful to the faculty at Courant for enlightening me on the subject, as well as for sponsoring HPC resources that made all the computation possible. In particular, I greatly appreciate Gordor Ritter, who has had several meaningful discussions with me and has provided tremendous support for this study. His prior research is also the one that gives me a lot of inspirations. In addition, I'd like to thank Oriol Vinyals, one of the leads on the AlphaStar project from Deepmind, for clearing many of my doubts about DRL.
\bigskip

\nocite{*}

\bibliographystyle{siamplain}

\end{document}